\documentstyle{mn}
\newif\ifAMStwofonts
\ifoldfss
  \ifCUPmtlplainloaded \else
    \NewTextAlphabet{textbfit} {cmbxti10} {}
    \NewTextAlphabet{textbfss} {cmssbx10} {}
    \NewMathAlphabet{mathbfit} {cmbxti10} {} 
    \NewMathAlphabet{mathbfss} {cmssbx10} {} 
  \fi
  \ifAMStwofonts
    \ifCUPmtlplainloaded \else
      \NewSymbolFont{upmath} {eurm10}
      \NewSymbolFont{AMSa} {msam10}
      \NewMathSymbol{\upi}     {0}{upmath}{19}
      \NewMathSymbol{\umu}     {0}{upmath}{16}
      \NewMathSymbol{\upartial}{0}{upmath}{40}
      \NewMathSymbol{\leqslant}{3}{AMSa}{36}
      \NewMathSymbol{\geqslant}{3}{AMSa}{3E}

    \fi
  \fi
\fi 

\ifnfssone
  \newmathalphabet{\mathit}
  \addtoversion{normal}{\mathit}{cmr}{m}{it}
  \addtoversion{bold}{\mathit}{cmr}{bx}{it}
  \newmathalphabet{\mathbfit} 
  \addtoversion{normal}{\mathbfit}{cmr}{bx}{it}
  \addtoversion{bold}{\mathbfit}{cmr}{bx}{it}
  \newmathalphabet{\mathbfss} 
  \addtoversion{normal}{\mathbfss}{cmss}{bx}{n}
  \addtoversion{bold}{\mathbfss}{cmss}{bx}{n}
  \ifAMStwofonts
    \ifCUPmtlplainloaded \else
      \UseAMStwoboldmath
      \makeatletter
      \new@mathgroup\upmath@group
      \define@mathgroup\mv@normal\upmath@group{eur}{m}{n}
      \define@mathgroup\mv@bold\upmath@group{eur}{b}{n}
      \edef\UPM{\hexnumber\upmath@group}
      \new@mathgroup\amsa@group
      \define@mathgroup\mv@normal\amsa@group{msa}{m}{n}
      \define@mathgroup\mv@bold\amsa@group{msa}{m}{n}
      \edef\AMSa{\hexnumber\amsa@group}
      \makeatother
      \mathchardef\upi="0\UPM19
      \mathchardef\umu="0\UPM16
      \mathchardef\upartial="0\UPM40
      \mathchardef\leqslant="3\AMSa36
      \mathchardef\geqslant="3\AMSa3E
    \fi
  \fi
\fi 

\ifnfsstwo
  \DeclareMathAlphabet{\mathbfit}{OT1}{cmr}{bx}{it}
  \SetMathAlphabet\mathbfit{bold}{OT1}{cmr}{bx}{it}
  \DeclareMathAlphabet{\mathbfss}{OT1}{cmss}{bx}{n}
  \SetMathAlphabet\mathbfss{bold}{OT1}{cmss}{bx}{n}
  \ifAMStwofonts
    \ifCUPmtlplainloaded \else
      \DeclareSymbolFont{UPM}{U}{eur}{m}{n}
      \SetSymbolFont{UPM}{bold}{U}{eur}{b}{n}
      \DeclareSymbolFont{AMSa}{U}{msa}{m}{n}
      \DeclareMathSymbol{\upi}{0}{UPM}{"19}
      \DeclareMathSymbol{\umu}{0}{UPM}{"16}
      \DeclareMathSymbol{\upartial}{0}{UPM}{"40}
      \DeclareMathSymbol{\leqslant}{3}{AMSa}{"36}
      \DeclareMathSymbol{\geqslant}{3}{AMSa}{"3E}
    \fi
  \fi
\fi 

\ifCUPmtlplainloaded \else
  \ifAMStwofonts \else 
    \def\upi{\pi}
    \def\umu{\mu}
    \def\upartial{\partial}
  \fi
\fi

\title[]
  {Clusters AgeS Experiment.
RR Lyr Variables in the Globular Cluster NGC 6362}
\author[A. Olech et al.]
  {A.~Olech,$^{1}$~  J.~Kaluzny,$^{2}$~
  I.B.~Thompson,$^3$~ W.~Pych,$^1$~ W.~Krzemi\'nski,$^{2,3}$~ 
\newauthor
and A.~Schwarzenberg-Czerny$^{2,4}$\\
  $^1$Warsaw University Observatory, Al. Ujazdowskie 4,
00-478 Warsaw, Poland (olech,pych@sirius.astrouw.edu.pl)\\
  $^2$Copernicus Astronomical Center,
ul. Bartycka 18, 00-716 Warsaw, Poland (jka,wk,alex@camk.edu.pl)\\
  $^3$Carnegie Observatories, 813 Santa Barbara Street,
Pasadena,CA 91101, USA (ian@ociw.edu)\\
  $^4$Astronomical Observatory of Adam Mickiewicz University,
ul. Sloneczna 36, 61-286 Poznan, Poland\\
}
\date{Accepted 2000 .............. Received 2000 .............}
\pagerange{\pageref{firstpage}--\pageref{lastpage}}
\pubyear{1999}

\def\LaTeX{L\kern-.36em\raise.3ex\hbox{a}\kern-.15em
    T\kern-.1667em\lower.7ex\hbox{E}\kern-.125emX}

\begin{document}

\label{firstpage}

\maketitle

\begin{abstract} 

We present $V$ and $B$-band CCD photometry of 35 RR Lyr stars in the
southern hemisphere globular cluster NGC 6362. Fourier decomposition of
the light curves was used to estimate the basic properties of these
variables. From the analysis of the RRc stars we obtain a mean mass of
$M=0.531\pm0.014~M_\odot$, luminosity $\log L/L_\odot=1.656\pm0.006$,
effective temperature $T_{eff}=7429\pm20$, and helium abundance
$Y=0.292\pm0.002$. The mean values of the absolute magnitude,
metallicity (on Jurcsik's scale) and effective temperature for 14 RRab
stars with "regular" light curves are: $M_V=0.86\pm0.01$, ${\rm
[Fe/H]}=-0.93\pm0.04$ and $T_{eff}=6555\pm25$, respectively.
From the $B-V$ colors, periods and metallicities of the RRab stars
we estimate the color excess for NGC 6362 to be equal to
$E(B-V)=0.08\pm0.01$. Using this value we derive the colors of the blue
and red edges of the instability strip in NGC 6362 to be
$(B-V)^{BE}_0=0.17$ and $(B-V)^{RE}_0=0.43$.
From the relations between the Fourier coefficients of RRab and RRc stars
and their absolute magnitudes we estimate the apparent distance modulus to NGC
6362 to be $(m-M)_V=14.46\pm0.10$. From the mean value of $\log
L/L_\odot$ of the  RRc stars we obtain $14.59\pm0.03$.

The $V$-band light curves of three of the RRc stars exhibit changes in
amplitude of over 0.1 mag on the time scale of around one week. Near
the radial 1st overtone frequency we find one or two peaks, which
strongly suggests that these variables belong to the newly identified
group of non-radial pulsating RR Lyr stars.

\end{abstract}


\section{Introduction}

The Clusters AgeS Experiment (CASE) is a long term project with the main
goal  of determining accurate ages and distances of globular clusters by
using observations of detached eclipsing binaries (Paczy\'nski 1997).
NGC~6362 was selected as one of the targets of our ongoing photometric
survey  aimed at identification of binaries suitable for detailed
analysis (Kaluzny et al. 1999; Thompson et al. 1999).  As a byproduct we
obtain time series photometry of RR~Lyr stars located in the surveyed
clusters (Olech et al. 1999a, 1999b, Kaluzny et al. 2000)  

NGC 6362 (C$1726-670$) is one of the southern hemisphere Oosterhoff type
I globular clusters. The low central concentration of the cluster and
its proximity enable the measurement of  reliable photometry for the
majority of its stars.  The Third Catalogue of Variable Stars in
Globular Clusters (Sawyer Hogg 1973) contains a list of 33 variables in
NGC 6362. An updated list of these stars with more precise period
determinations was published by Clement et al. (1995). Three objects
from this list (V4, V9 and V32) were noted as non variable stars. The
nature of V1 was listed as unknown and the other objects were classified
as RR Lyr stars.

Recently Mazur et al. (1999) published results of a CCD search for
short-period variables in NGC 6362. They discovered 19 new candidates,
five of which were cluster RR Lyr stars, four were SX Phe stars and
eight were eclipsing binaries.

The metallicity of NGC 6362 is moderately high and relatively well
established. On the scale of Zinn and West (1984), [Fe/H] is equal to
$-1.08$. Recently, Carretta and Gratton (1997)  obtained ${\rm
[Fe/H]=-0.96}$ from an analysis of echelle spectra of cluster giants.

Alcaino and Liller (1986) obtain a distance modulus $(m-M)_V=14.74$ and
a reddening $E(B-V)=0.10$. Recently, Piotto et al. (1999) obtained
$(m-M)_V=14.67\pm0.20$ and $E(B-V)=0.06\pm0.03$ from an analysis of
Hubble Space Telescope data.

\section{Observations and Data Reduction}

Time series photometry of NGC~6362 was obtained during the interval
1999 April 17 -- August 22 with the 1.0-m Swope telescope at Las Campanas
Observatory.  We used the backside-illuminated SITE3 $2048\times 4096$
CCD, with a pixel scale 0.435 arcsec/pixel. A $2048\times3150$
subsection of the CCD  was used giving a field of view of $14.8\times22.8$
arcmin$^{2}$.  Photometry presented in this paper is based on 1088
$V$-band images and 259 $B$-band images which were collected on 42
nights. Exposure times were adjusted on every night -- depending on the
seeing -- to assure unsaturated images for stars from the horizontal
branch of the cluster. For the $V$-band these ranged from 180 to 300
seconds. 
 
Instrumental photometry was extracted using DoPHOT (Schechter, Mateo
and Saha 1993). We used DoPHOT in the fixed-position mode, with the
stellar positions measured on "template" images constructed by
averaging the 3 best exposures obtained for each filter. The
instrumental photometry was transformed to the standard $BV$ system
using observations of standard stars from Landolt (1992). We estimate
that systematic errors of the zero points of the final $BV$ photometry
do not exceed 0.03 mag.

\section{Results}

Our search for pulsating variable stars in NGC 6362 has identified 35
RR Lyrae stars. All these variables were previously known (Clement et
al., 1995, Mazur et al., 1999, we use the nomenclature from these
papers). The periods $P$, $V$ filter peak-to-peak amplitudes $A_V$,
intensity averaged $<V>$ and $<B>$ magnitudes and types of these stars
are presented in Table 1. The phased light curves of these stars are
shown in Fig. 1.  Crowding prevented the classification of V1 and V32
by Clement et al. and Mazur et al. The light curves in Fig. 1 indicate
that both are RRab stars.

\begin{table}
\caption{Elements of the pulsating variables in NGC 6362}
\begin{tabular}{rccccl}
\hline
\hline
Star & $P$ [days] & $A_V$ & $<V>$ & $<B>$ & Type \\
\hline
\hline
V1  & 0.504800  & 1.09  & 15.274 & 15.657 &  RRab\\
V2  & 0.488973  & 1.27  & 15.243 & 15.595 &  RRab\\
V3  & 0.447110  & 1.14  & 15.266 & 15.598 &  RRab\\
V5  & 0.520730  & 0.74  & 15.332 & 15.741 &  RRab\\
V6  & 0.262703  & 0.47  & 15.288 & 15.547 &  RRc\\
V7  & 0.521590  & 1.03  & 15.276 & 15.664 &  RRab\\
V8  & 0.381480  & 0.44  & 15.087 & 15.362 &  RRc\\
V10  & 0.265675  & 0.47  & 15.296 & 15.552 &  RRc\\
V11  & 0.288790  & 0.50  & 15.219 & 15.475 &  RRc\\
V12  & 0.532905  & 1.02  & 15.230 & 15.581 &  RRab\\
V13  & 0.579987  & 1.00  & 15.211 & 15.573 &  RRab\\
V14  & 0.246210  & 0.34  & 15.337 & 15.588 &  RRc\\
V15  & 0.279940  & 0.43  & 15.221 & 15.540 &  RRc\\
V16  & 0.525668  & 1.08  & 15.299 & 15.706 &  RRab\\
V17  & 0.314600  & 0.42  & 15.308 & 15.639 &  RRc\\
V18  & 0.512876  & 0.96  & 15.306 & 15.706 &  RRab\\
V19  & 0.594506  & 0.59  & 15.318 & 15.794 &  RRab\\
V20  & 0.698330  & 0.38  & 15.206 & 15.713 &  RRab\\
V21  & 0.281390  & 0.49  & 15.267 & 15.558 &  RRc\\
V22  & 0.266840  & 0.51  & 15.317 & 15.582 &  RRc\\
V23  & 0.275110  & 0.49  & 15.303 & 15.574 &  RRc\\
V24  & 0.329360  & 0.46  & 15.109 & 15.432 &  RRc\\
V25  & 0.455891  & 1.28  & 15.407 & 15.722 &  RRab\\
V26  & 0.602182  & 0.61  & 15.327 & 15.789 &  RRab\\
V27  & 0.278120  & 0.51  & 15.280 & 15.545 &  RRc\\
V28  & 0.358470  & 0.44  & 15.074 & 15.406 &  RRc\\
V29  & 0.647743  & 0.42  & 15.218 & 15.688 &  RRab\\
V30  & 0.613372  & 0.94  & 15.137 & 15.542 &  RRab\\
V31  & 0.600230  & 0.60  & 15.294 & 15.739 &  RRab\\
V32  & 0.497203  & 1.14  & 15.277 & 15.659 &  RRab\\
V33  & 0.306420  & 0.42  & 15.318 & 15.636 &  RRc\\
V34  & 0.494370  & 0.68  & 15.326 & 15.723 &  RRab\\
V35  & 0.290790  & 0.43  & 15.254 & 15.576 &  RRc\\
V36  & 0.310090  & 0.38  & 15.107 & 15.475 &  RRc\\
V37  & 0.255062  & 0.28  & 15.273 & 15.478 &  RRc\\
\hline
\hline
\end{tabular}
\end{table}

We compared our photometry with the observations of Mazur et al.
(1999).  For 18 RR Lyr stars in common, the mean difference $<V>_{\rm
our}-<V>_{\rm Mazur}$ is equal to $0.001\pm0.006$.

In Fig. 2 we present a color-magnitude diagram for RR Lyr variables in
NGC 6362. Open circles denote the RRc stars, and open triangles
correspond to the RRab variables.

The periods of the cluster RRc stars are between 0.2462 and 0.3815 days with
a mean value of 0.295 days. The periods of the RRab variables are between
0.4471 and 0.6983 days with a mean value of 0.538 days. These values
are in very good agreement with the mean values of periods of RRc and
RRab stars in NGC 6362 of 0.296 and 0.556 days, respectively, obtained
by Clement et al. (1995). These mean values are also similar to those
obtained for the Oosterhoff type I clusters NGC 6171 and NGC 6732 with
characteristics resembling NGC 6362 (Sandage 1993).

We fitted our $V$-band light curves with Fourier sine series of the
form:
\begin{equation}
V=A_0+\sum^{12}_{j=1}A_j\cdot\sin(j\omega t + \phi_j)
\end{equation}

\noindent where $\omega=2\pi/P$. To find the values of $\omega$, $A_j$
and $\phi_j$ we employed the method developed by Schwarzenberg-Czerny
(1997) and Schwarzenberg-Czerny and Kaluzny (1998). To demonstrate
precision of our light curves,  in Fig. 3 we plot the observed light
curve of RRab variable V25 together with the Fourier sine series with
$j=15$. The observational points are plotted using grey open circles of
diameter 0.06 mag, equal to $10$ standard deviations. The solid black
line corresponds to the fit based on equation (1). Table 2 presents the
amplitudes $A_j$ and phases $\phi_j$ and their errors.  For RRab stars
with amplitudes of 1.25 mag we can determine values of $A_j$ up to
$j=14$ with relative error smaller than 10\%. The relative error of
$A_{15}$ is equal to 14\%.

\begin{table}
\caption{Fourier Elements of the RRab variable V25 from NGC 6362}
\begin{tabular}{|cccccc|}
\hline
\hline
$A_1$ & $\sigma A_1$ & $A_2$ & $\sigma A_2$ & $A_3$ & $\sigma A_3$ \\
0.4401 & 0.0003 & 0.2118 & 0.0003 & 0.1576 & 0.0003 \\
\hline
$A_4$ & $\sigma A_4$ & $A_5$ & $\sigma A_5$ & $A_6$ & $\sigma A_6$ \\
0.1010 & 0.0003 & 0.0737 & 0.0003 & 0.0508 & 0.0003 \\
\hline
$A_7$ & $\sigma A_7$ & $A_8$ & $\sigma A_8$ & $A_9$ & $\sigma A_9$ \\
0.0320 & 0.0003 & 0.0211 & 0.003 & 0.0173 & 0.0003 \\
\hline
$A_{10}$ & $\sigma A_{10}$ & $A_{11}$ & $\sigma A_{11}$ & $A_{12}$ & $\sigma A_{12}$ \\
0.0137 & 0.0003 & 0.0097 & 0.0003 & 0.0059 & 0.0003 \\
\hline
$A_{13}$ & $\sigma A_{13}$ & $A_{14}$ & $\sigma A_{14}$ & $A_{15}$ & $\sigma A_{15}$\\
0.0035 & 0.0003 & 0.0029 & 0.0003 & 0.0022 & 0.0003\\
\hline
\hline
$\phi_1$ & $\sigma\phi_1$ & $\phi_2$ & $\sigma\phi_2$ &$\phi_3$ & $\sigma\phi_3$ \\
5.169 & 0.001 & 0.041 & 0.002 & 1.506 & 0.002 \\
\hline
$\phi_4$ & $\sigma\phi_4$ & $\phi_5$ & $\sigma\phi_5$ &$\phi_6$ & $\sigma\phi_6$ \\
3.007 & 0.003 & 4.538 & 0.005 & 6.129 & 0.007 \\
\hline
$\phi_7$ & $\sigma\phi_7$ & $\phi_8$ & $\sigma\phi_8$ &$\phi_9$ & $\sigma\phi_9$ \\
1.327 & 0.010 & 2.674 & 0.016 & 4.080 & 0.019 \\
\hline
$\phi_{10}$ & $\sigma\phi_{10}$ & $\phi_{11}$ & $\sigma\phi_{11}$ &$\phi_{12}$ & $\sigma\phi_{12}$ \\
5.556 & 0.024 & 0.875 & 0.034 & 2.362 & 0.055 \\
\hline
$\phi_{13}$ & $\sigma\phi_{13}$ & $\phi_{14}$ & $\sigma\phi_{14}$ &$\phi_{15}$ & $\sigma\phi_{15}$ \\
3.683 & 0.094 & 5.198 & 0.117 & 0.100 & 0.155 \\
\hline
\hline
\end{tabular}
\end{table}

The values of the peak-to-peak amplitudes $A_V$ and periods $P$ presented in
Table 1 are used to plot the period-amplitude ($\log_{10}P-A_V$) diagram
shown in Fig. 4. Open circles denote RRc stars, filled triangles RRab
stars with $D_m<3$ (for definition of $D_m$ see section 3.3) and open
triangles RRab variables with $D_m>3$. The solid line represents a linear
fit to RRab variables in M3 (Kaluzny et al. 1998) another Oosterhoff
type I globular cluster.

Clement and Shelton (1999) have suggested that for RRab stars the $V$
amplitude at a given period is not a function of metal abundance, but
rather only a function of the Oosterhoff type. We see in Fig. 4 that
the majority of our RRab variables lay below the mean line for
Oosterhoff type I clusters, as defined by the RRab stars in the
globular cluster M3. A similar result is also found by Kaluzny et al.
(2000) for the cluster M5 and by Wehlau, Slawson, and Nemec (1999) for
the cluster NGC~7006. These results suggest that while there is a
clear correlation of Oosterhoff type with metallicity (see Table 4),
within at least the Oosterhoff type I clusters there is an additional
dependence of $V$ amplitude at a given period upon metallicity, with
the more metal rich stars having lower $V$ amplitudes.

\subsection{Fourier Analysis of the RRc Stars}

Simon and Clement (1993) showed that the
Fourier decomposition of the light curves of RRc variables is a very useful
technique for estimating the physical parameters of these stars. The
equations of Simon and Clement (1993) are:
\begin{equation}
\log M = 0.52\log P_1 - 0.11\phi^*_{31} + 0.39
\end{equation}
 
\begin{equation}
\log L = 1.04\log P_1 - 0.058\phi^*_{31} + 2.41
\end{equation}
 
\begin{equation}
\log T_{eff} = 3.265 - 0.3026\log P_1 - 0.1777\log M + 0.2402\log L
\end{equation}
 
\begin{equation}
\log Y = -20.26 + 4.935\log T_{eff} - 0.2638\log M+ 0.3318\log L
\end{equation}

\noindent where $M$ is the mass of the star in solar units, $P_1$ is the
first overtone pulsation period in days, $L$ is the luminosity in  solar
units, $T_{eff}$ is the effective temperature in Kelvin, $Y$ is the
relative helium abundance and $\phi^*_{31}=\phi^*_3-3\phi^*_1$. The
phases marked by asterisks are obtained from a cosine Fourier series
(used by Simon and Clement 1993) and differ from our phases which were
obtained from a sine series (cf. equation 1). For $\phi_{31}$ we have
$\phi_{31}=\phi^*_{31}+\pi$.

In addition, using the formula of Kov\'acs (1998) we may estimate the 
absolute magnitude $M_V^{Ko}$ of RRc stars directly:

\begin{equation}
M_V^{Ko}=1.261-0.961 P_1-0.004\phi_{21}-4.447 A_4
\end{equation}

\begin{table*}
\caption{Parameters for the RRc Variables in NGC 6362}
  \begin{tabular}{lrrrrrrrrrr}
\hline
\hline
Star & $A_0$ & $A_1$ & $\phi_{21}$ & $\phi_{31}$ & $M$ & $\log L$ &
$T_{eff}$ & $Y$ & $M_V^{Ko}$ & $M_V$\\
\hline
\hline
V6 & 15.304 & 0.244 & 3.166 & 6.476 & 0.526 &1.613 &7544 &0.305 &0.825 & 0.753\\
   &  $\pm$0.000 &  $\pm$0.003 & $\pm$0.105 & $\pm$0.190 & $\pm$0.025 &$\pm$0.011 &  $\pm$18 &$\pm$0.006 &$\pm$0.014& \\
V8 & 15.096 & 0.217 & 4.182 & 7.778 & 0.460 &1.706 &7267 &0.282 &0.657 & 0.520\\
   &  $\pm$0.000 &  $\pm$0.001 & $\pm$0.084 & $\pm$0.032 & $\pm$0.004 &$\pm$0.002 &   $\pm$2 &$\pm$0.001 &$\pm$0.006& \\
V10& 15.311 & 0.242 & 3.151 & 6.154 & 0.574 &1.637 &7500 &0.295 &0.818& 0.693\\
   &  $\pm$0.000 &  $\pm$0.002 & $\pm$0.070 & $\pm$0.136 & $\pm$0.020 &$\pm$0.008 &  $\pm$13 &$\pm$0.004 &$\pm$0.009& \\
V11& 15.234 & 0.254 & 3.315 & 6.614 & 0.534 &1.648 &7454 &0.294 &0.762& 0.665\\
   &  $\pm$0.000 &  $\pm$0.001 & $\pm$0.017 & $\pm$0.027 & $\pm$0.004 &$\pm$0.002 &   $\pm$2 &$\pm$0.001 &$\pm$0.005& \\
V14& 15.343 & 0.169 &3.047 & 6.090 & 0.561 &1.606 &7577 &0.305 &0.881& 0.770\\
   &  $\pm$0.000 &  $\pm$0.000 & $\pm$0.015 & $\pm$0.068 & $\pm$0.010 &$\pm$0.004 &   $\pm$6 &$\pm$0.002 &$\pm$0.001& \\
V15& 15.234 & 0.219 & 2.959 & 6.201 & 0.583 &1.657 &7448 &0.289 &0.804& 0.643\\
   &  $\pm$0.000 &  $\pm$0.000 & $\pm$0.016 & $\pm$0.027 & $\pm$0.004 &$\pm$0.002 &   $\pm$2 &$\pm$0.001 &$\pm$0.005& \\
V17& 15.316 & 0.210 & 3.172 & 7.267 & 0.473 &1.648 &7425 &0.298 &0.788& 0.665\\
   &  $\pm$0.000 &  $\pm$0.001 & $\pm$0.033 & $\pm$0.044 & $\pm$0.005 &$\pm$0.003 &   $\pm$4 &$\pm$0.001 &$\pm$0.005& \\
V21& 15.281 & 0.253 & 3.173 & 6.398 & 0.557 &1.648 &7461 &0.293 &0.775& 0.665\\
   &  $\pm$0.000 &  $\pm$0.000 & $\pm$0.009 & $\pm$0.013 & $\pm$0.002 &$\pm$0.001 &   $\pm$1 &$\pm$0.000 &$\pm$0.000& \\
V22& 15.317 & 0.254 & 3.156 & 5.972 & 0.603 &1.649 &7478 &0.290 &0.790& 0.663\\
   &  $\pm$0.000 &  $\pm$0.000 & $\pm$0.007 & $\pm$0.017 & $\pm$0.003 &$\pm$0.001 &   $\pm$1 &$\pm$0.000 &$\pm$0.000& \\
V23& 15.305 & 0.253 & 3.134 & 6.086 & 0.595 &1.656 &7456 &0.288 &0.779& 0.645\\
   &  $\pm$0.000 &  $\pm$0.000 & $\pm$0.013 & $\pm$0.021 & $\pm$0.003 &$\pm$0.001 &   $\pm$1 &$\pm$0.001 &$\pm$0.001& \\
V24& 15.121 & 0.231 & 3.329 & 7.303 & 0.480 &1.667 &7379 &0.292 &0.740& 0.618\\
   &  $\pm$0.000 &  $\pm$0.001 & $\pm$0.054 & $\pm$0.064 & $\pm$0.008 &$\pm$0.004 &   $\pm$6 &$\pm$0.002 &$\pm$0.005& \\
V27& 15.296 & 0.261 & 3.091 & 6.195 & 0.582 &1.655 &7455 &0.289 &0.773& 0.648\\
   &  $\pm$0.000 &  $\pm$0.001 & $\pm$0.014 & $\pm$0.022 & $\pm$0.003 &$\pm$0.001 &   $\pm$2 &$\pm$0.001 &$\pm$0.004& \\
V28& 15.085 & 0.221 & 3.684 & 7.562 & 0.470 &1.690 &7313 &0.286 &0.692& 0.560 \\
   &  $\pm$0.000 &  $\pm$0.001 & $\pm$0.045 & $\pm$0.031 & $\pm$0.004 &$\pm$0.002 &   $\pm$2 &$\pm$0.001 &$\pm$0.005& \\
V33& 15.328 & 0.215 & 3.161 & 7.022 & 0.497 &1.651 &7429 &0.296 &0.796& 0.658\\
   &  $\pm$0.000 &  $\pm$0.000 & $\pm$0.020 & $\pm$0.029 & $\pm$0.004 &$\pm$0.002 &   $\pm$2 &$\pm$0.001 &$\pm$0.001& \\
V35& 15.267 & 0.217 & 3.038 & 6.525 & 0.548 &1.656 &7438 &0.291 &0.803& 0.645\\
   &  $\pm$0.000 &  $\pm$0.001 & $\pm$0.021 & $\pm$0.033 & $\pm$0.005 &$\pm$0.002 &   $\pm$3 &$\pm$0.001 &$\pm$0.005& \\
V36& 15.113 & 0.189 & 3.178 & 7.113 & 0.488 &1.651 &7425 &0.296 &0.788& 0.658\\
   &  $\pm$0.000 &  $\pm$0.001 & $\pm$0.044 & $\pm$0.058 & $\pm$0.007 &$\pm$0.003 &   $\pm$5 &$\pm$0.002 &$\pm$0.005& \\
V37& 15.277 & 0.138 & 2.034 & 4.061 & 0.956 &1.740 &7343 &0.252 &0.904& 0.435\\
   &  $\pm$0.000 &  $\pm$0.003 & $\pm$0.199 & $\pm$5.529 & $\pm$1.339 &$\pm$0.321 & $\pm$520 &$\pm$0.010 &$\pm$0.016& \\
\hline
\hline
\end{tabular}
\end{table*}

From Eqs. (2)--(6) we compute masses, luminosities,
effective temperatures, relative helium abundances and absolute
magnitudes of RRc stars in NGC 6362. These are presented in Table 3
together with the values of $A_0$, $A_1$, $\phi_{21}$ and $\phi_{31}$.
The errors presented in Table 3 are calculated from the formal errors of
the Fourier coefficients using the error propagation law.

We exclude from our sample variables V6, V10 and V37 due to their
irregular light curves and large errors of $\phi_{31}$ (see section
3.2). The mean values of the mass, luminosity, effective temperature and
helium abundance for the remaining 14 RRc stars are
$0.531\pm0.014~M_\odot$, $\log L/L_\odot=1.656\pm0.006$,
$T_{eff}=7429\pm20$, and $Y=0.292\pm0.002$, respectively.
The mean $\log P_1$ is equal to $-0.525\pm0.014$ and the mean
$\phi^*_{31}$ to $3.582\pm0.162$.

\begin{table*}
\caption{Mean parameters for RRc stars in
globular clusters after Kaluzny et al. (1998, 2000).}
\begin{tabular}{|l|c|c|c|c|c|c|c|}
\hline
\hline
Cluster & Oosterhoff & [Fe/H] & No. of & mean & mean & mean & mean \\
        &   type     &        & stars  & mass & $\log L$ & $T_{eff}$ &$Y$ \\
\hline
\hline
NGC 6171 & I & -0.68 & 6 & 0.53 & 1.65 & 7447 & 0.29\\
NGC 6362 & I & -1.08 &14 & 0.53 & 1.66 & 7429 & 0.29\\
M5       & I & -1.25 &14 & 0.54 & 1.69 & 7353 & 0.28\\      
M3       & I & -1.47 & 5 & 0.59 & 1.71 & 7315 & 0.27\\
M9       & II& -1.72 & 1 & 0.60 & 1.72 & 7299 & 0.27\\
M55      & II& -1.90 & 5 & 0.53 & 1.75 & 7193 & 0.27\\
NGC 2298 & II& -1.90 & 2 & 0.59 & 1.75 & 7200 & 0.26\\
M68      & II& -2.03 &16 & 0.70 & 1.79 & 7145 & 0.25\\
M15      & II& -2.28 & 6 & 0.73 & 1.80 & 7136 & 0.25\\
\hline
\hline
\end{tabular}
\end{table*}

By comparing the physical parameters of the RRc stars from different
clusters Clement and Shelton (1997) revealed several correlations.
Specifically, an increasing mean mass corresponds to an increasing
luminosity and a decreasing effective temperature and relative helium
abundance. It is clearly visible from Table 4 taken from Kaluzny et al.
(1998, 2000) that NGC 6362 fits this sequence rather well.

\subsection{Non-radial Pulsations of RRc Stars}

Three of the RRc variables presented in Fig. 1, namely, V37, V6 and
V10, show clear modulation of their light curves. The amplitudes and
shapes of the light curves change on the time scale of days. This is
not the Blazhko effect, which is observed mainly in RRab stars as a
modulation of the amplitude on a time scale of tens of days. The other
possible explanation is that these stars are bimodal pulsators (RRd
stars) with pulsation in both the fundamental mode and the 1st
overtone. To check this possibility we calculated the power spectra of
these variables. The power spectrum of the light curve of V37 is
presented in the upper panel of Fig. 5. The highest peak is detected at
a frequency of $f_1=3.920615$ cyc/day ($P_1=0.2550062^d$). The light
curve of this star phased with $P_1$ is shown in the third plot in the
upper panel. One can see that the light curve is noisy. The
power spectrum shows a secondary peak at frequency $f_2=3.984365$
($P_2=0.250981^d$). We prewhitened our original light curve removing
the main period $P_1$ with its harmonics. The power spectrum of the
prewhitened light curve is shown in the middle panel of Fig. 5. There
is no trace of the main frequency $f_1$ and the highest and only
significant peak is $f_2$. The third plot in the middle panel presents
the prewhitened light curve phased with the period $P_2$. Again we
prewhitened this light curve now removing period $P_2$ with its
harmonics, and the resulting power spectrum shows no further
significant frequencies (see the lower panel of Fig. 5). We conclude
that both frequencies correspond to real modes of pulsation. For
ordinary double mode pulsators the ratio of the 1st overtone period to
the fundamental period is approximately 0.745. In our case the ratio
$P_2/P_1$ is equal to 0.98 and thus it is clear that if $P_1$
corresponds to the radial 1st overtone pulsations then $P_2$ in V 37
corresponds to a non-radial mode.

Another variable with a "noisy" light curve is V6. The power spectrum of
its raw light curve is presented in the upper panel of Fig. 6. The
highest peak has frequency $f_1=3.806580$ cyc/day ($P_1=0.262703^d$).
The 
power spectrum reveals the presence of other components. We prewhitened
the light curve of V6 using
$P_1$, and the power spectrum of this prewhitened light curve is presented in
the second panel of Fig. 6. There are two high peaks with frequencies of
$f_2=3.740961$ ($P_2=0.267311^d$) and $f_3=3.871003$~cyc/day
($P_3=0.258331^d$) in this plot. We again prewhitened our light curve
using the period $P_2$ and its harmonics, and the power spectrum of
the twice prewhitened light curve is presented in the third panel
of Fig. 6. The only peak that is left is $f_3$. After removing $P_3$ with
its harmonics we obtained a power spectrum clean of any significant
features, presented in the lower panel in Fig. 6. 

The frequencies $f_3$ and $f_2$ lay symmetrically above and below $f_1$,
with $f_3-f_1$ = 0.0644 and $f_1-f_2$ = 0.0656.  This symmetry suggests
that $P_2$ and $P_3$ are signatures of the amplitude modulation with a
period equal to $1/(f_1-f_2)=1/(f_3-f_1)\approx15$ days. This resembles
the Blazhko effect. Although the Blazhko effect is common among RRab
stars, it is seldomly observed in RRc stars. In particular, in a sample
of 46 stars with known Blazhko periods there are only three RRc stars,
and the shortest Blazhko period is 10.9 days (Smith 1995). However, the
difference $(f_1-f_2)-(f_3-f_1)=0.0012$ is large compared to the
accuracy of our frequency determinations. Since the separate light
curves phased with periods $P_1$, $P_2$ and $P_3$ shown in Fig. 6
display stable amplitudes, they could correspond to pulsations in three
different modes of which two are non-radial, similar to the one
non-radial mode in V37.

A similar analysis was performed for variable V10 with the results
presented in Fig. 7. In this case we also detected three periods. The
most significant is $P_1=0.265675^d$ ($f_1=3.763997$) and two other are
$P_2=0.257573^d$ ($f_2=3.882395$) and $P_3=0.274262^d$
($f_3=3.646148$). Again the differences of frequencies $f_2-f_1$ and
$f_1-f_3$ are similar and equal to 0.1184 and 0.1178, respectively. The
corresponding beat period $1/(f_2-f_1)=1/(f_1-f_3) is \approx8.5$ days. If
we assume that the amplitude modulation is the Blazhko effect, 
corresponding Blazhko period would be the shortest known! For V10 the
difference $(f_2-f_1)-(f_1-f_3)$ is equal to 0.0006 and, given  the stable
light curves in Fig. 7, we suggest that the 
periods $P_2$ and $P_3$ are real and correspond to non-radial
pulsations.

Non-radial pulsations in RR Lyr stars were first discovered by Olech et
al. (1999a) in three RRc variables in the globular cluster M55. This
discovery was quickly followed by  Olech et al. (1999b)
who found one non-radial pulsating RRc star in M5, by Kov\'acs et al. (2000) 
who reported a few such RRc stars
in the  MACHO data,  and by Moskalik (2000) who
discovered six non-radial RRc and RRab pulsators in the OGLE data.

A theoretical investigation of non-radial pulsations in RR Lyr stars
was presented by Dziembowski and Cassisi (1999). They showed that the
largest growth rates are seen for radial modes and strongly trapped
unstable modes with large values of $l$. The latter do not explain our
observations because it is difficult to detect  pulsations with such
large values of $l$ ($l=7-10$) in photometric data. The third group of
modes with quite large growth rates are non-radial pulsations with
$l=1$ laying in the vicinity of the unstable 1st overtone mode (see
Fig. 2 of Dziembowski and Cassisi, 1999). The most unstable $l=1$ modes
always have high frequencies (short periods) which is in good agreement
with our observations. In NGC 6362 there is only one monoperiodic RRc
star with a period shorter than the periods of the RRc stars with
non-radial pulsations. The remaining 13 monoperiodic RRc stars have
longer periods.  A similar situation was observed by Olech et al.
(1999a) in M55. Among nine RRc stars there is only one star with a
period shorter than the period of the three non-radially pulsating
variables.

Dziembowski and Cassisi (1999) showed also that the relative frequency
separation $\Delta f/f$ is largest for the $l=1$ modes in the vicinity
of the 1st overtone pulsations. For the shortest periods $\Delta f/f$ is
around 0.01 which is in good agreement with our observations which give
0.016, 0.017 and 0.031 for V37, V6 and V10, respectively.

The $B-V$ color range for RRc variables in NGC 6362 is between 0.21 and
0.33. The colors of  V37, V6 and V10 are 0.21, 0.26 and 0.26,
respectively, placing these stars near the blue edge of the instability
strip. The same result was obtained by Olech at al (1999a) who found
three non-radial pulsating RRc stars in M55 with $B-V$ colors between
0.31 and 0.32. The $B-V$ colors of the RRc stars in M55 are between 0.31 and
0.39.

\subsection{Fourier Analysis of the RRab Variables}

Recently Kov\'acs and Jurcsik (1996, 1997) and Jurcsik (1998) have
derived formulae that estimate the absolute magnitudes, metallicities,
intrinsic colors and temperatures of RRab stars from the periods,
amplitudes and phases as measured from a Fourier analysis of the light
curves.
These equations are:

\begin{equation}
{\rm [Fe/H]} = -5.038 - 5.394 P_0 + 1.345\phi_{31}
\end{equation}
\begin{equation}
M_V = 1.221 - 1.396 P_0 - 0.477A_1 + 0.103\phi_{31}
\end{equation}
\begin{equation}
V_0-K_0=1.585 + 1.257 P_0 - 0.273A_1 - 0.234\phi_{31} + 0.062\phi_{41}
\end{equation}
\begin{equation}
\log T_{eff} = 3.9291 - 0.1112(V_0-K_0) - 0.0032{\rm [Fe/H]}
\end{equation}
 
\noindent where $\phi_{41}=\phi_4-4\phi_1$ (cf. equation 1).
 
These equations are valid only for RRab stars with regular light
curves, i.e. variables with a deviation parameter $D_m$  smaller than 3
(see Kov\'acs and Kanbur 1998). In our sample, only three variables do not
satisfy this condition. Table 5 summarizes the results obtained from
equations (7)-(10). The mean values of the absolute magnitude,
metallicity and effective temperature for 14 stars with $D_m<3$ are
$M_V=0.86\pm0.01$, ${\rm [Fe/H]}=-0.93\pm0.04$ and $T_{eff}=6555\pm25$,
respectively.

\begin{table*}
\caption{Parameters for the RRab Variables in NGC 6362}
  \begin{tabular}{lrrrrrrrrr}
\hline
\hline
Star & P[days] & $A_0$ & $A_1$ & $\phi_{31}$ & $\phi_{41}$ & $M_V$ & [Fe/H] &
$T_{eff}$ & $D_m$ \\
\hline
\hline
V1  & 0.504800 &15.321  & 0.376 & 5.015 & 1.412  & 0.859 & $-$1.015  & 6572 & 1.64\\
    &          & $\pm$0.001  & $\pm$0.001  & $\pm$0.009  & $\pm$0.013  & $\pm$0.084  & $\pm$0.022\\
V2  & 0.488973 &15.309  & 0.422 & 5.040 & 1.357  & 0.861 & $-$0.897 & 6637 & 1.83\\
    &          & $\pm$0.001  & $\pm$0.001  & $\pm$0.008  & $\pm$0.012  & $\pm$0.084  & $\pm$0.022\\
V3  & 0.447110 &15.322  & 0.449 & 5.013 & 1.293  & 0.904 & $-$0.707  & 6726 & 2.38\\
    &          & $\pm$0.001  & $\pm$0.002  & $\pm$0.020  & $\pm$0.038  & $\pm$0.082  & $\pm$0.035\\
V5  & 0.520730 &15.354  & 0.290 & 5.220 & 1.554  & 0.899 & $-$0.825  & 6556 &  2.49\\
    &          & $\pm$0.002  & $\pm$0.003  & $\pm$0.061  & $\pm$0.089  & $\pm$0.090  & $\pm$0.084\\
V7  & 0.521590 &15.323  & 0.361 & 5.113 & 1.580  & 0.852 & $-$0.974  & 6549 & 1.58\\
    &          & $\pm$0.001  & $\pm$0.001  & $\pm$0.017  & $\pm$0.024  & $\pm$0.087  & $\pm$0.028\\
V12 & 0.532905 &15.278  & 0.361 & 5.063 & 1.487  & 0.831 & $-$1.103  & 6521 & 1.49\\
    &          & $\pm$0.002  & $\pm$0.002  & $\pm$0.027  & $\pm$0.040  & $\pm$0.086  & $\pm$0.040\\
V13 & 0.579987 &15.253  & 0.362 & 5.241 & 1.864  & 0.784 & $-$1.118  & 6454 & 1.76\\
    &          & $\pm$0.001  & $\pm$0.001  & $\pm$0.018  & $\pm$0.026  & $\pm$0.090  & $\pm$0.031\\
V16 & 0.525668 &15.346  & 0.359 & 5.105 & 1.547  & 0.847 & $-$1.007  & 6541  & 0.80\\
    &          & $\pm$0.000  & $\pm$0.000  & $\pm$0.004  & $\pm$0.006  & $\pm$0.087  & $\pm$0.018\\
V18 & 0.512876 &15.351  & 0.344 & 5.141 & 1.541  & 0.876 & $-$0.890  & 6571 & 1.50\\
    &          & $\pm$0.002  & $\pm$0.003  & $\pm$0.040  & $\pm$0.055  & $\pm$0.087  & $\pm$0.056\\
V19 & 0.594506 &15.332  & 0.220 & 5.522 & 2.378  & 0.860 & $-$0.818  & 6402 &15.26\\
    &          & $\pm$0.000  & $\pm$0.000  & $\pm$0.010  & $\pm$0.020  & $\pm$0.096  & $\pm$0.033\\
V20 & 0.698330 &15.212  & 0.163 & 5.974 & 3.459  & 0.790 & $-$0.769  & 6227 & 4.71\\
    &          & $\pm$0.000  & $\pm$0.000  & $\pm$0.019  & $\pm$0.058  & 0.107  & $\pm$0.060\\
V25 & 0.455891 &15.478  & 0.440 & 4.849 & 1.181  & 0.879 & $-$0.976  & 6662 & 1.51\\
    &          & $\pm$0.000  & $\pm$0.000  & $\pm$0.004  & $\pm$0.005  & $\pm$0.080  & $\pm$0.024\\
V26 & 0.602182 &15.343  & 0.227 & 5.575  & 2.507  & 0.852 & $-$0.788  & 6395 &  2.08\\
    &          & $\pm$0.000  & $\pm$0.001  & $\pm$0.011  & $\pm$0.024  & $\pm$0.097  & $\pm$0.035\\
V29 & 0.647743 &15.226  & 0.184 & 5.738  & 2.314  & 0.826 & $-$0.814  & 6366 &15.76\\
    &          & $\pm$0.001  & $\pm$0.001  & $\pm$0.052  & 0.114  & 0.102  & $\pm$0.081\\
V30 & 0.613372 &15.170  & 0.316 & 5.379 & 2.075  & 0.773 & $-$1.111  & 6396 &  2.12\\
    &          & $\pm$0.001  & $\pm$0.001  & $\pm$0.012  & $\pm$0.018  & $\pm$0.094  & $\pm$0.029\\
V31 & 0.600230 &15.308  & 0.223 & 5.617 & 2.421  & 0.861 & $-$0.720  & 6419 & 13.89\\
    &          & $\pm$0.001  & $\pm$0.001  & $\pm$0.019  & $\pm$0.036  & $\pm$0.098  & $\pm$0.043\\
V32 & 0.497203 &15.330  & 0.380 & 5.040 & 1.422  & 0.870 & $-$0.941  & 6596 & 1.50\\
    &          & $\pm$0.001  & $\pm$0.001  & $\pm$0.012  & $\pm$0.018  & $\pm$0.085  & $\pm$0.024\\
V34 & 0.494370 &15.349  & 0.263 & 5.273 & 1.658  & 0.954 & $-$0.612  & 6599 &  2.64\\
    &          & $\pm$0.001  & $\pm$0.002  & $\pm$0.031  & $\pm$0.071  & $\pm$0.090  & $\pm$0.047\\
\hline
\hline
\end{tabular} 
\end{table*}

In their original paper Jurcsik and Kov\'acs (1996) introduced the
$D_m$ parameter to distinguish  RRab variables with regular light curves
($D_m<3$) from those with irregular light curves, i.e.  those showing
Blazhko behavior ($D_m>3$). In our sample of RRab stars V5, V7, V12 and
V13 exhibit clear amplitude modulations. Surprisingly, the $D_m$
parameter for these stars is less than 3.  This may be a result of the
large number of observational points in our light curves.  The Blazhko
effect is then averaged over many cycles and the fit given in  equation
(1) is not able to detect the modulation since the light curve seems to
be regular.

The estimated mean metallicity ${\rm [Fe/H]}=-0.93\pm0.04$ is in 
very good agreement with the estimates of Zinn and West (1984) ${\rm
[Fe/H]}=-1.08$ and Carretta and Gratton (1997) ${\rm [Fe/H]}=-0.96$. 
However, the metallicity computed from
equation (7) is on the scale of Jurcsik (1995) which is connected with
scale of Zinn and West by the formula:

\begin{equation}
{\rm [Fe/H]_{Jurcsik}} = 1.431{\rm [Fe/H]_{ZW}}+0.880
\end{equation}

\noindent If we take ${\rm [Fe/H]}=-1.08$ from Zinn and West (1984),
the equation (11) gives [Fe/H] = $-0.67$ on the Jurcsik (1995) scale.
The reason for this discrepancy is unknown. Previous estimates of
[Fe/H] using equation (7) have produced results in broad agreement with
other estimates (see Kaluzny et al., 1998, 2000, Olech et al.
1999a).

\subsection{Variable V18}

The light curve of RRab variable V18 shown in Fig. 1 is quite noisy. The
scatter is observed not only around maximum as in Blazhko variables
(e.g. V5, V7, V12, V13) but along the whole light curve. This may indicate
that this variable pulsates in two modes. The non-radial pulsations
discovered in RRc stars (this paper, Olech et al., 1999a) were also observed in RRab
stars (Moskalik, 2000) and the presence of non-radial modes in V18
could explain the scatter in the light curve.

Inspection of the raw light curve suggests an amplitude modulation of
0.1 mag to 0.2 mag on a time scale of around 40 days,
perhaps indicating a beating effect. In our opinion there
is another possible explanation.

The power spectrum shown in Fig. 8 displays a maximum frequency at
$f_0=1.94979$~c/d which corresponds to $P_0=0.512876^d$. There are also
many aliases of $f_0$ placed at $f_0\pm1$, $f_0+2$ and $f_0+3$. The AoV
spectrum for three harmonics also shows a clear variation at $f_0$ with
weak peaks at $f_0\pm1$, $f_0/2$, $f_0/3$, $f_0/4$ and $(f_0+1)/2$.

Both these periodograms are in agreement with a single frequency $f_0$
but this does not explain the scatter in the light curve of V18. One can
notice that the doubled period ($2\cdot P_0=1.026^d$) differs only 0.026
days from one day and thus at each 38 days around midnight we observe
only the flat part of the light curve which is close to the mean brightness
of the star. After 20 days around midnight we observe the largest
change of the amplitude. Thus the modulation of the amplitude might not
be real but might be caused by observing different parts of the light
curve at each night.

The prewhitening of the raw light curve with $f_0$ and five harmonics
gives a noisy light curve with a standard deviation of 0.07 mag, 
much larger than the photometric errors. The power spectrum of the
prewhitened light curve shows many peaks with the highest at $f_0$ and
$f_0/2$ with amplitudes of approximately 0.03 mag.

We conclude that there is no evidence for  multiperiodic
behavior of V18. It is possible that the enhanced scatter detected in
the light curve of this star is partially caused by the Blazhko effect but 
this
is difficult to check due to the proximity of $P_0$ to half a day.

\subsection{Reddening and the instability strip of NGC 6362}

The reddening of RRab variables can be calculated from the metallicity,
expressed in terms of $\Delta S$ (Preston 1959). From Blanco (1992) we
have the relation:
 
\begin{equation}
E(B-V)=<B-V>_{\Phi(0.5-0.8)}+0.01222\Delta S 
\end{equation}
 
\noindent $- 0.00045(\Delta S)^2 - 0.185P -0.356$
\vspace{0.2cm}
 
\noindent where $<B-V>_{\Phi(0.5-0.8)}$ is the observed mean color in
the 0.5-0.8 phase interval and $P$ is the fundamental period. Based on
the globular cluster metallicity scale adopted by Zinn and West (1984)
Suntzeff et al. (1991) derived the following $\Delta S - {\rm [Fe/H]}$
relation:
 
\begin{equation}
{\rm [Fe/H]} = -0.408 - 0.158\Delta S
\end{equation}
 
\noindent Using the equation (13) and adopting ${\rm [Fe/H]=-1.0}$ for
NGC 6362 we obtain $\Delta S=3.75$.

The average value of $E(B-V)$ calculated in this way for our 18 NGC 6362
RRab stars is $E(B-V)=0.08\pm0.01$. This value is consistent with
recent determination of Piotto et al. (1999) who obtained
$E(B-V)=0.06\pm0.03$, slightly smaller than the value of Alcaino and
Liller (1986) who obtained $E(B-V)=0.10$.
 
Knowing the colors of RR Lyr variables and the reddening we are able to
estimate the boundaries of the instability strip in NGC 6362. Our
coolest variable is V20 with $<B>-<V>$ equal to 0.51 and  $T_{eff}$
computed from equation (10) equal to 6227~K. The
hottest variable is V14 with $<B>-<V>$ equal to 0.25 and  $T_{eff}$
computed from equation (4) equal to 7577~K (we exclude V37 with
$<B>-<V>=0.21$ due to its noisy light curve). Adopting a
color excess of $E(B-V)=0.08$, we measure the blue and red edges of 
the instability
strip in NGC 6362 to be at $(B-V)^{BE}_0=0.17$ and 
$(B-V)^{RE}_0=0.43$. These values are consistent with estimates made for
other globular clusters (Smith 1995).

\subsection{Distance Modulus to NGC 6362}

Piotto et al. (1999) estimate the apparent distance modulus to NGC 6362
to be $(m-M)_V=14.67\pm0.20$, based on a measured mean magnitude of the
horizontal branch (HB) and the average of a number of calibrations of the
relation between the absolute magnitude of the zero age horizontal
branch and the cluster metallicity.

A comparison of the color magnitude diagram of NGC 6362 presented in
Fig. 2 with that of Piotto et al. (1999) shows a small but significant
shift in $V$ magnitude. Piotto et al. (1999) present a histogram of the
distribution in $V$ for the HB stars in the color
interval $0.5<B-V<0.7$. From this histogram they obtained a mean visual
magnitude for the HB of $V_{\rm HB}=15.45\pm0.03$. Fig. 9 shows similar
histograms (we decided to increase the range of colors to
$0.0<B-V<0.85$) for all color magnitude diagrams of the cluster found
in the literature. The panels correspond to the work of (from up to
bottom) Alcaino and Liller (1986), Mazur et al (1999), Piotto et al.
(1999) and to this work. All histograms, except that of Piotto et al.,
show that the mean $V$ magnitude of the HB is equal to 15.33.
The 0.12 mag difference between the Piotto et al. estimate and the
other data sets in Fig. 9 suggests that the value of apparent distance
modulus to the NGC 6362 from the Piotto et al. work should be
$(m-M)_V=14.55\pm0.20$ instead of $(m-M)_V=14.67\pm0.20$.

Knowing the values of $A_0$ and $M_V^{Ko}$ for RRc stars (see Table 3)
we can compute the apparent distance modulus to the NGC 6362. It is
equal to $(m-M)_V=14.46\pm0.02$,  quite consistent  with our corrected
value of Piotto's determination. On the other hand one should remember
that the equation for $M_V^{Ko}$ is calibrated using Baade-Wesselink
luminosities, which give faint absolute magnitudes for RR Lyr stars. We
can also use the values of $\log L/L_\odot$ of RRc stars to compute the
distance modulus. The absolute magnitudes $M_V$ (presented in the last
column of Table 3) were calculated assuming a value of 4.79 for $M_{\rm
bol}$ of the Sun  and using a bolometric correction ${\rm BC}=0.06{\rm
[Fe/H]}+0.06$, adopted from Sandage and Cacciari (1990). The resulting
distance modulus is equal to $(m-M)_V=14.59\pm0.03$, in excellent
agreement  with the value presented above. This agreement suggests that
the theoretical models constructed by Simon and Clement (1993) produce
the correct values of luminosities of RRc stars.

A similar estimate can be made for the RRab stars.  Using the data
given  in Table 5 (14 RRab stars with $D_m<3$) we obtain
$(m-M)_V=14.46\pm0.10$. It is exactly the same value as we obtained
using Kov\'acs' expression for $M_V^{Ko}$ of RRc stars (see section
3.1). The agreement is expected since both equations (6) and (8) were
calibrated using Baade-Wesselink luminosities.

Table 6 summarizes the distance and reddening determinations for NGC 6362.

\begin{table}
\caption{The distance and reddening determinations for NGC 6362}
\begin{tabular}{|l|l|l|}
\hline
\hline
Distance modulus & Color excess & ~~~Source \\
~~~~$(m-M)_V$    & ~~$E(B-V)$  & \\
\hline
\hline
14.74          & 0.10          & Alcaino \& Liller (1986)\\
$14.67\pm0.20$ & $0.06\pm0.03$ & Piotto et al. (1999)\\
$14.55\pm0.20$ & ---           & Piotto corrected \\
$14.46\pm0.02$ & ---           & $M_V^{Ko}$ of RRc stars \\
$14.59\pm0.03$ & ---           & $\log L/L_\odot$ for RRc stars \\
$14.46\pm0.10$ & $0.08\pm0.01$ & RRab stars \\
\hline
\hline
\end{tabular}
\end{table}

The values of the distance modulus based on the mean magnitude of the
HB and that measured from $\log L/L_\odot$ of RRc stars are very
consistent. It is known that theoretical models of RR Lyr stars favor
the {\it long} distance scale and thus large values of distance
modulus. On the opposite side stand the statistical parallaxes and
Baade-Wesselink luminosities of RR Lyrae stars which give faint values
of absolute magnitudes of RR Lyr stars and thus  values of the distance
modulus which are smaller by 0.2-0.3 mag.

The final conclusion of this section is that the distance modulus to
NGC 6362 is around 14.45 mag if we prefer the {\it short} distance scale
and around 14.6 if we prefer the {\it long} distance scale.

\section{Conclusions}

We have presented a comprehensive study of 35 RR Lyr variables in the
globular cluster NGC 6362. Our $V$-band light curves contain over 1000
measurements and our $B$-band light curves around 260 measurements. The
periods of the cluster RRc stars are between 0.2462 and 0.3815 days
with a mean value of 0.295 days. The periods of the RRab variables
are between 0.4471 and 0.6983 days with a mean value of 0.538 days.

The period-amplitude ($\log_{10}P-A_V$) diagram for  NGC 6362 shows that
RRab variables from this cluster do not fit the $\log_{10}P-A_V$
relation for M3 another Oosterhoff type I globular cluster,
suggesting that  the $V$
amplitude for a given period is a function of metallicity rather than
Oosterhoff type alone.  

Fourier decomposition of the light curves was used to estimate the basic
properties of these variables. From the analysis of the RRc stars we
obtain a mean mass of $M=0.531\pm0.014~M_\odot$, luminosity $\log
L/L_\odot=1.656\pm0.006$, effective temperature $T_{eff}=7429\pm20$, and
helium abundance $Y=0.292\pm0.002$. The mean values of the absolute
magnitude, metallicity (Jurcsik's scale) and effective temperature for
14 RRab stars with "regular" light curves are: $M_V=0.86\pm0.01$, ${\rm
[Fe/H]}=-0.93\pm0.04$ and $T_{eff}=6555\pm25$, respectively.

Using the $B-V$ colors, periods and metallicities of the RRab stars, we
estimate the value of the color excess for NGC 6362 to be equal to
$E(B-V)=0.08\pm0.01$.  From this color excess we derive the unreddened
colors of the blue and red edges of the instability strip in NGC 6362
to be $(B-V)^{BE}_0=0.17$ and $(B-V)^{RE}_0=0.43$.

The apparent distance modulus of NGC 6362 calculated from the sample of
14 RRab stars is  $(m-M)_V=14.46\pm0.10$. An analysis of the RRc stars
based on  Kov\'acs' expression for $M_V^{Ko}$ leads to a value of
$(m-M)_V=14.46\pm0.02$. This agreement is expected since both
derivations rely on Baade-Wesselink luminosities, giving faint
magnitudes for RR Lyr stars and a {\it short} distance scale. On the
other hand, a larger value of the distance modulus is determined from
estimates of $\log L/L_\odot$ of the RRc stars --
$(m-M)_V=14.59\pm0.03$. This result is also expected because
theoretical models of RR Lyr stars produce large values of distance
moduli and prefer a {\it long} distance scale. These latter distance
determinations are consistent with the value of $(m-M)_V=14.55\pm0.20$
derived from the mean magnitude of the zero age horizontal branch.

Three of the RRc variables presented in Fig. 1 (V37, V6 and V10) show a
clear modulation of their light curves. The power spectra of the light
curves of these stars show the highest peaks for the 1st overtone
radial pulsations. In the case of V37 we detect another peak in the
vicinity of the main frequency which cannot be connected with radial
pulsations. In the case of V6 and V10 we detect two non-radial modes in
the vicinity of the main frequency. Our conclusion is that these
variables belong to the newly discovered group of non-radial pulsating
RR Lyr stars (Olech et al., 1999a, 1999b, Kov\'acs et al., 2000,
Moskalik, 2000). We have also noticed that the non-radial
pulsation RRc stars lay near the blue edge of the instability strip,
consistent with the results of Olech et al. (1999a) for RRc stars in
the globular cluster M55.

\section*{Acknowledgments} We would like to thank to Prof. W.
Dziembowski for helpful hints and comments.  AO and JK were supported
by the Polish Committee of Scientific Research through grant
2P03D$-$003$-$17 and by NSF grant AST$-$9528096 to Bohdan Paczy\'nski.
WP was supported by KBN grant 2P03D$-$020$-$15. IBT and WK acknowledge
support from NSF grant AST$-$9819787.

\clearpage
 
\noindent {\bf Fig. 1} Light curves of RR Lyr variable stars found in
NGC 6362. The stars are plotted by increasing period.
The upper light curve is in the$V$-band, and the lower in the $B$-band. 
The $B$ magnitude
is reduced by 0.2 mag for clarity.
 
\includegraphics{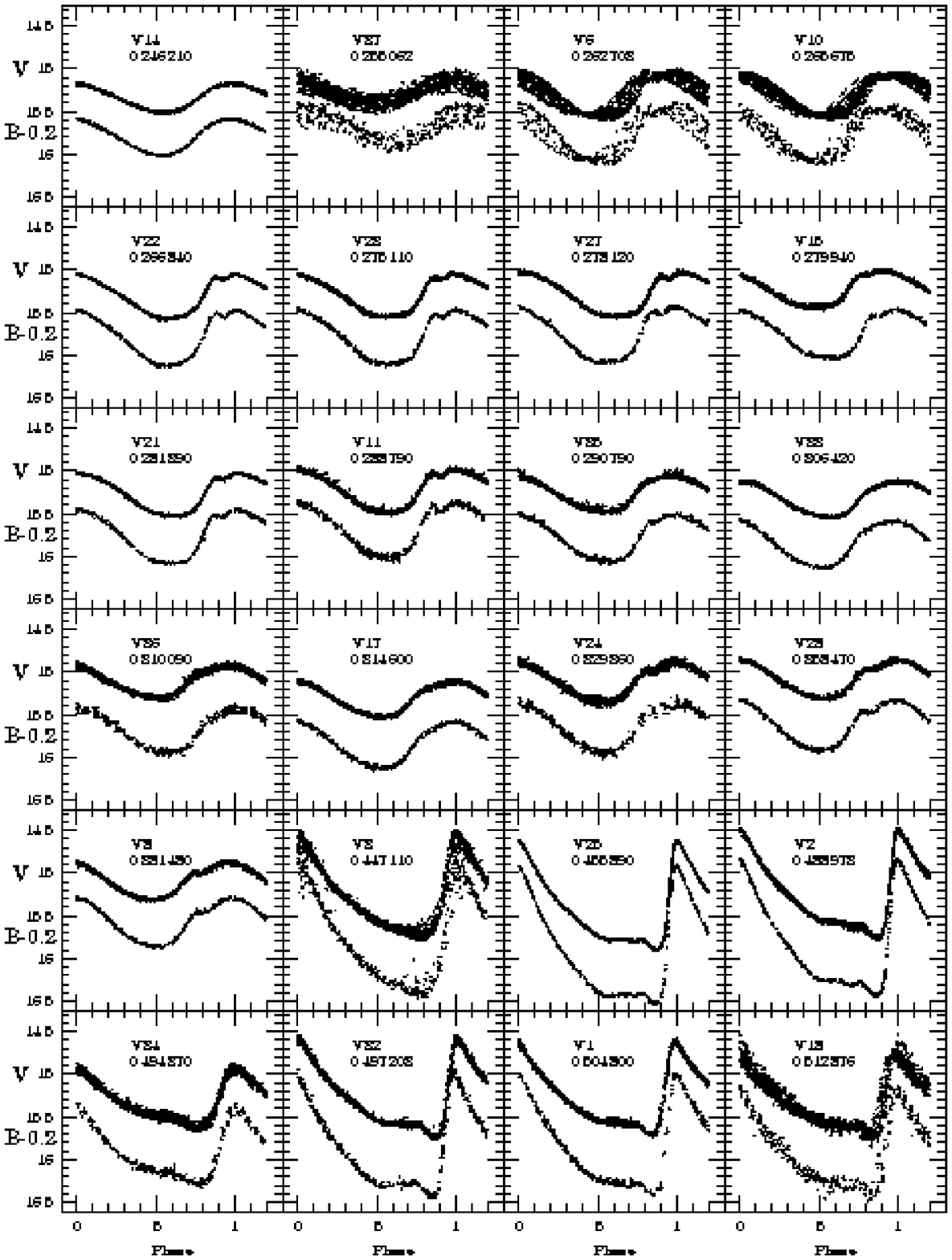}
 
\clearpage
 
\noindent {\bf Fig. 1} Continued.
 
\includegraphics{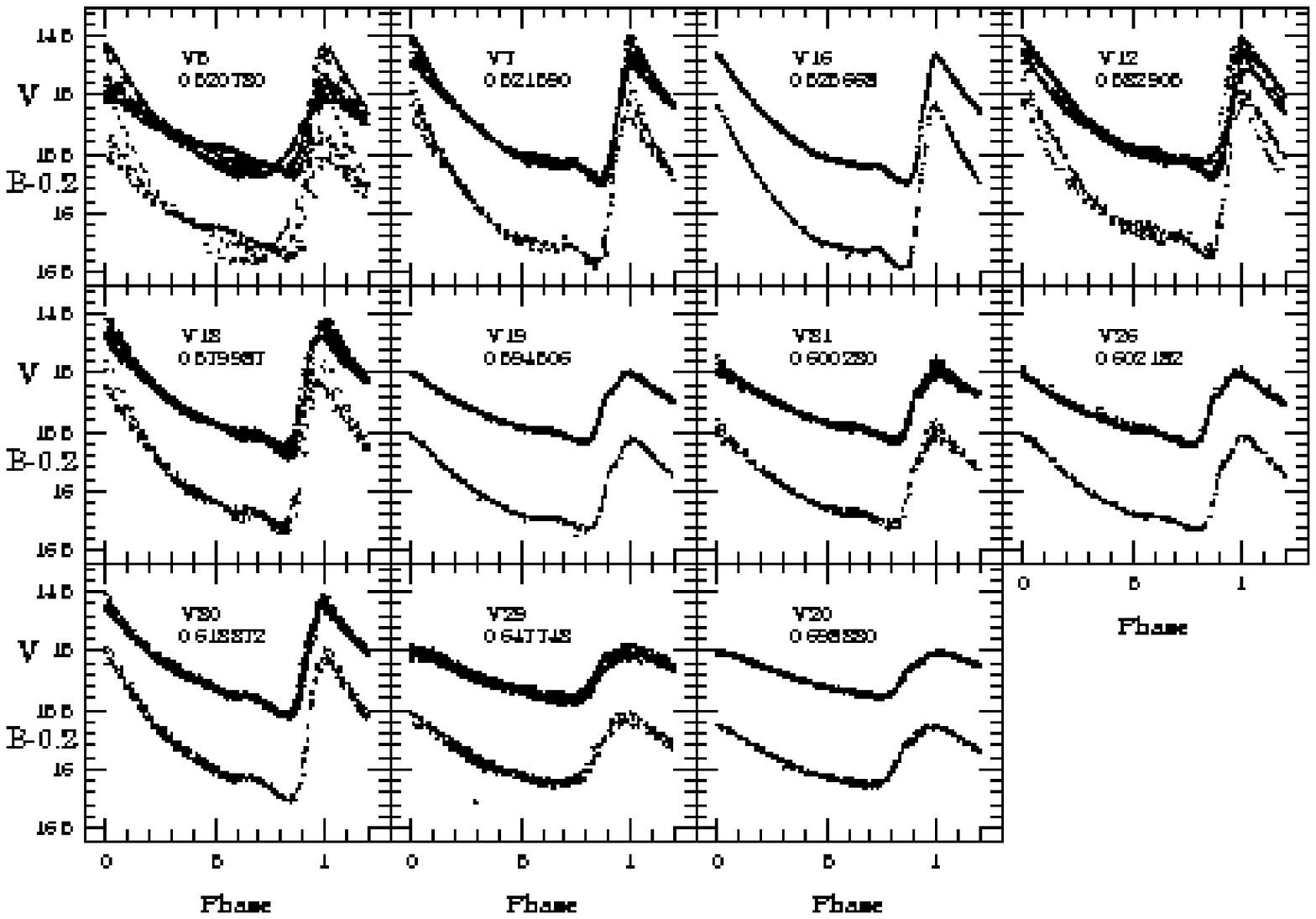}
 
\clearpage

\noindent {\bf Fig. 2} ~Color-magnitude diagram of NGC 6362. The open
triangles denote RRab stars and open circles correspond to RRc variables.

\includegraphics{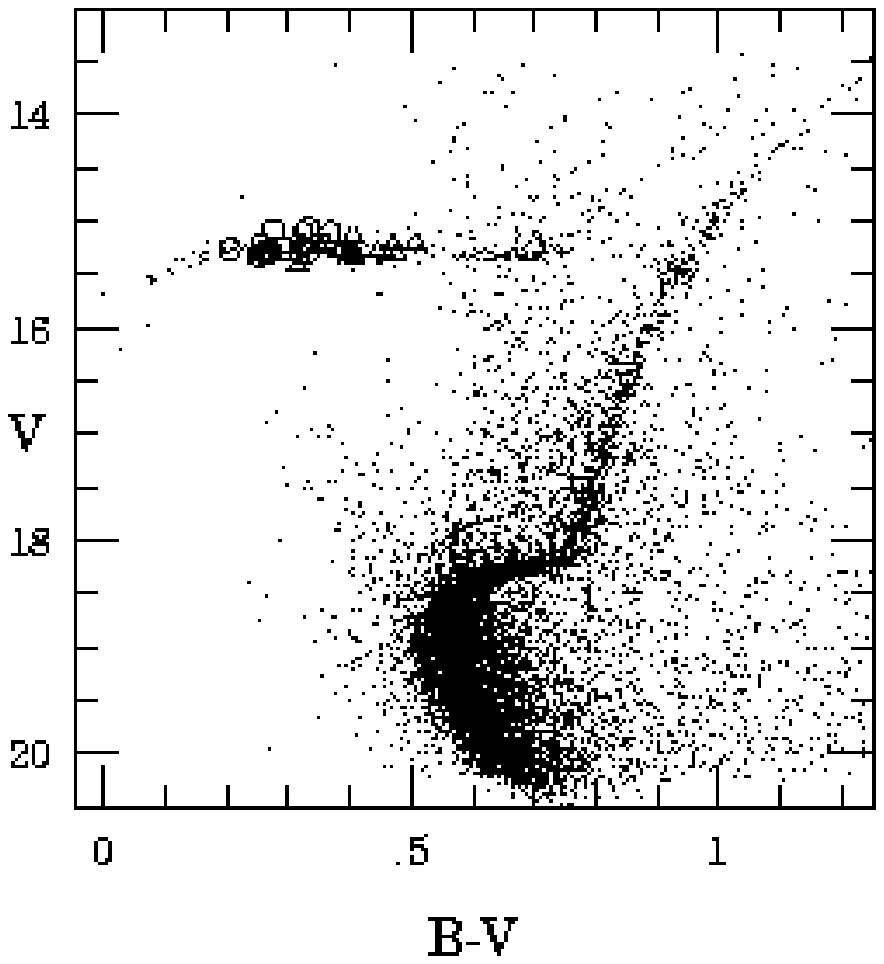}

\clearpage
 
\noindent {\bf Fig. 3} ~The phased light curve of variable V25. The
observational points are plotted with grey open circles. The size of
these circles is 10 times larger than the photometric errors. The solid
black line corresponds to the fit computed from equation (1).
\includegraphics{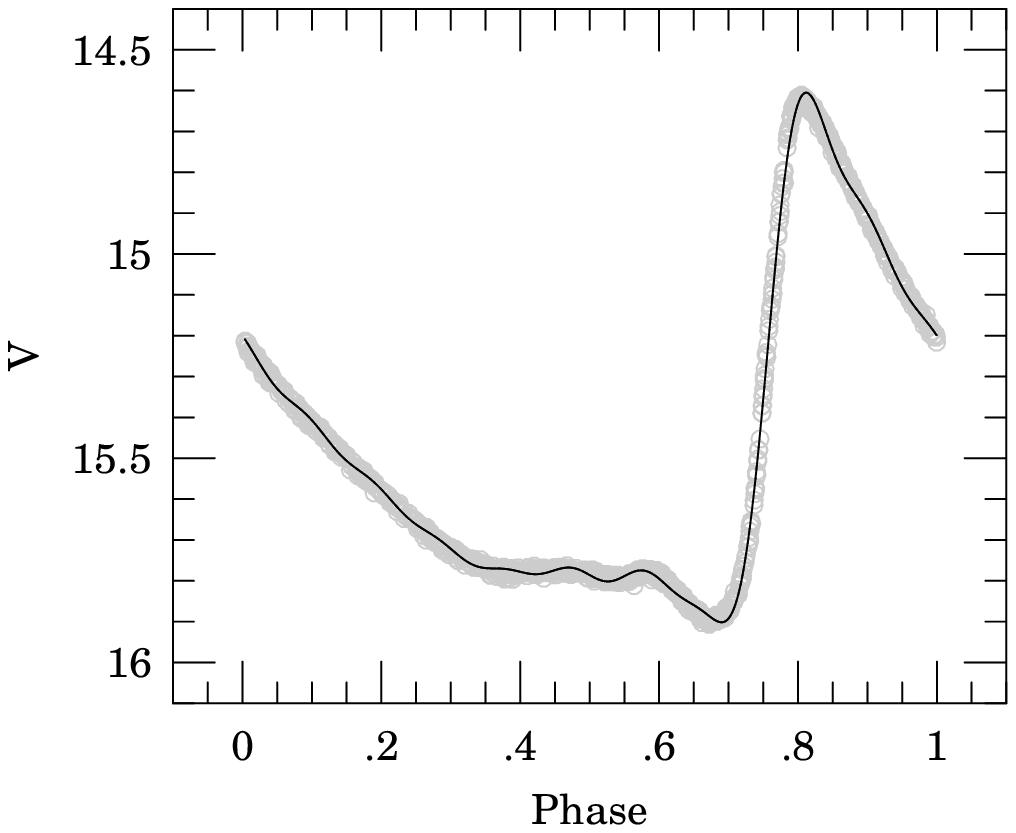}

\vspace{10cm}
 
\includegraphics{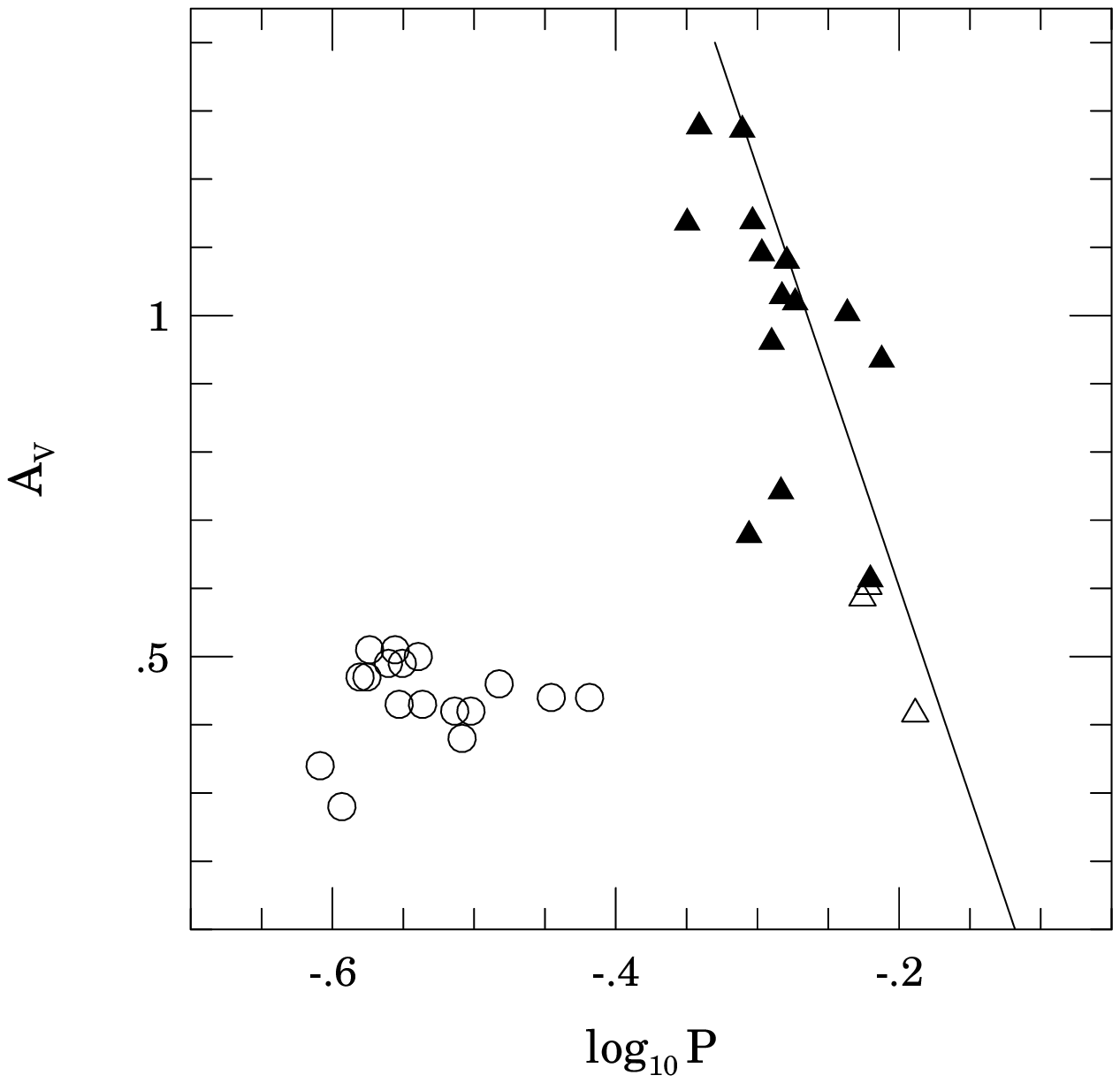}

\noindent {\bf Fig. 4} ~Period-amplitude diagram for RRab stars with
$D_m<3$ (solid triangles), RRab stars with $D_m>3$ (open triangles) and
RRc stars (open circles). The solid line represents a linear fit to RRab
variables in M3 (Kaluzny et al. 1998)

\clearpage
 
\noindent {\bf Fig. 5} ~The power spectrum of the real (top),
prewhitened (middle) and two times prewhitened (bottom) light curve of
variable V37. The prewhitened light curves phased with periods $P_1$ and
$P_2$ are shown on the right side.
\includegraphics{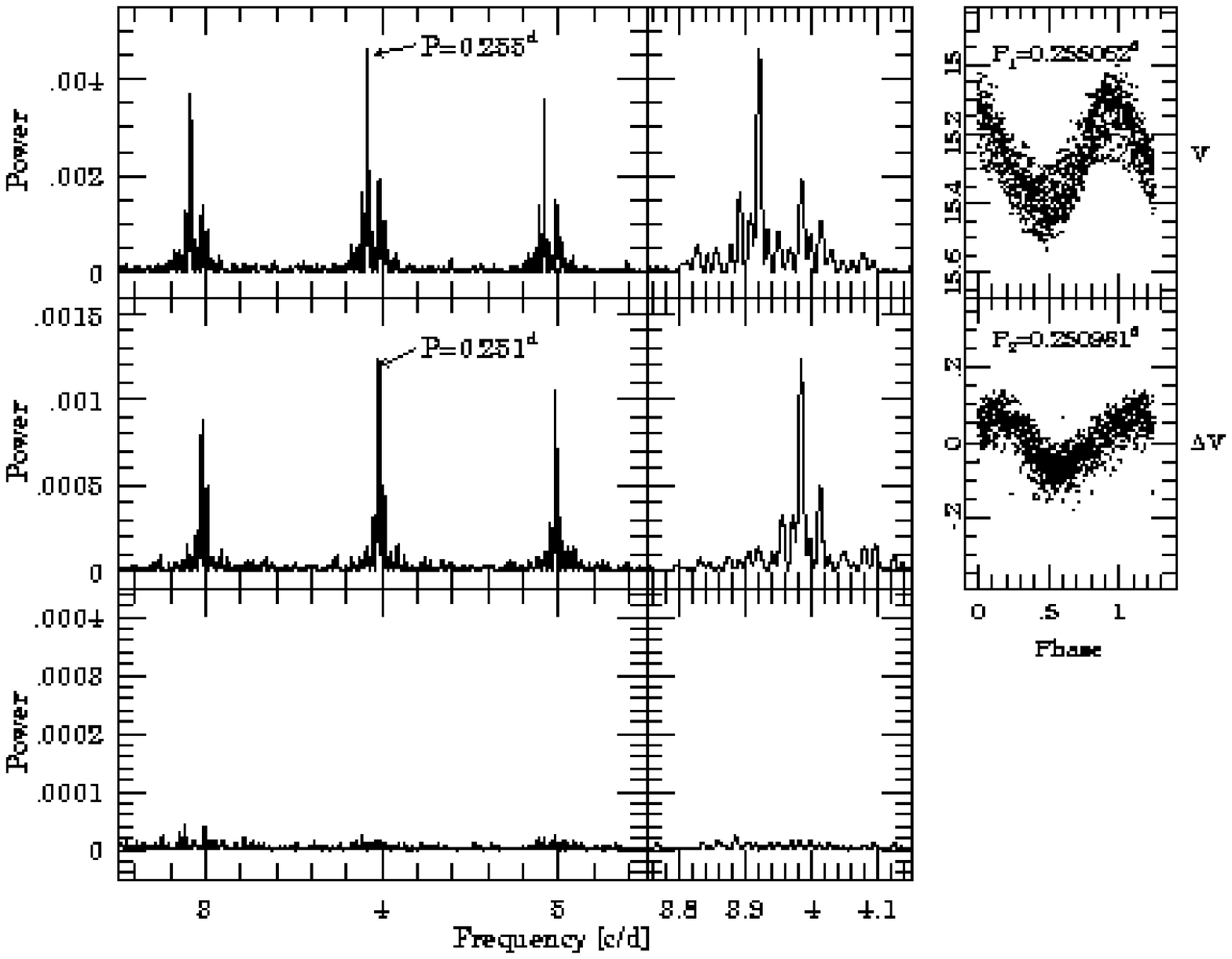}

\clearpage
 
\noindent {\bf Fig. 6} ~The power spectrum of the real (top),
prewhitened (upper middle), two times prewhitened (lower middle) and
three times prewhitened (bottom) light curve of
variable V6. The prewhitened light curves phased with periods $P_1$, $P_2$ and
$P_3$ are shown on the right side.
\includegraphics{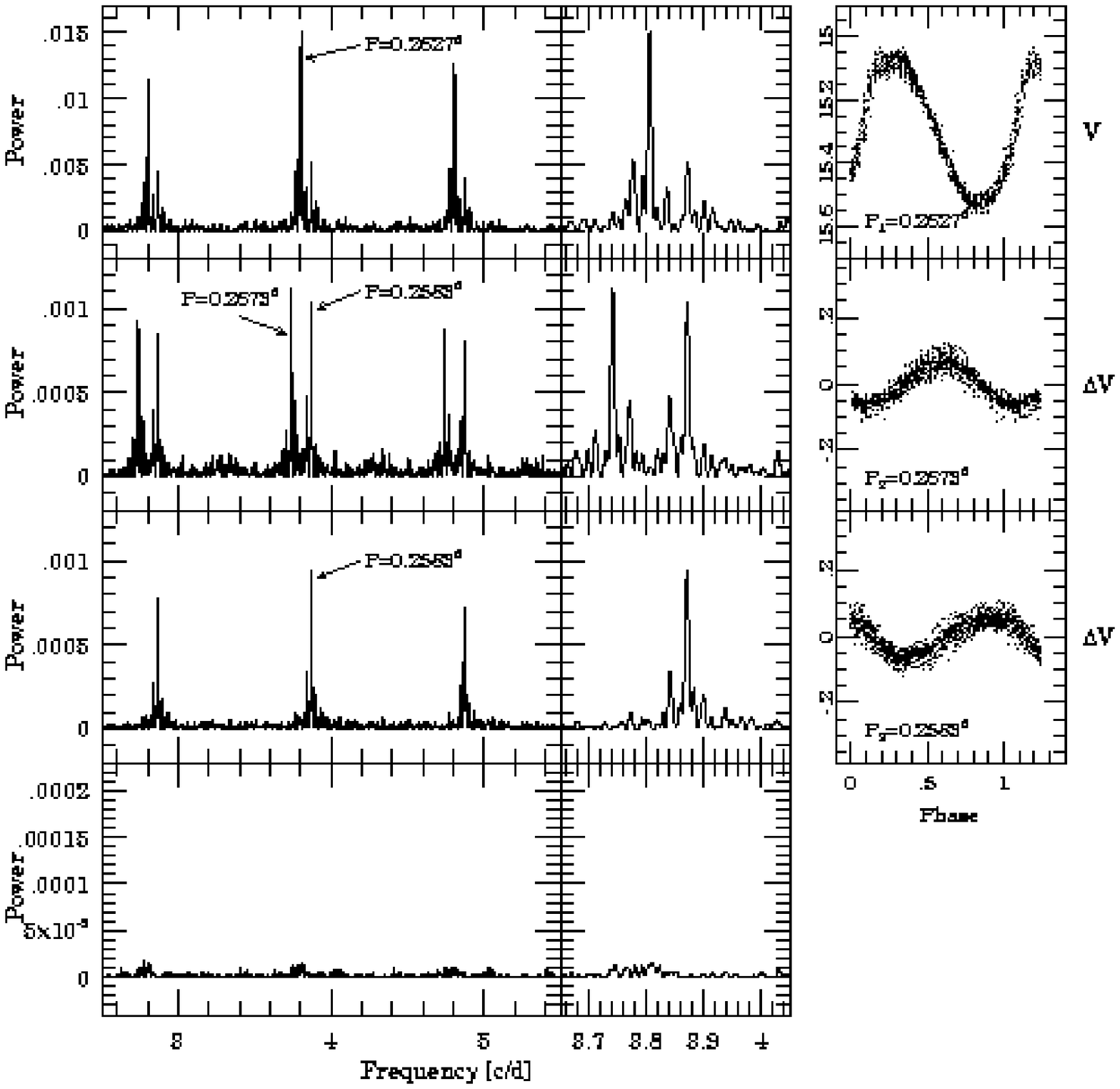}

\clearpage
 
\noindent {\bf Fig. 7} ~The power spectrum of the real (top),
prewhitened (upper middle), two times prewhitened (lower middle) and
three times prewhitened (bottom) light curve of
variable V10. The prewhitened light curves phased with periods $P_1$,
$P_2$ and $P_3$ are shown on the right side.
\includegraphics{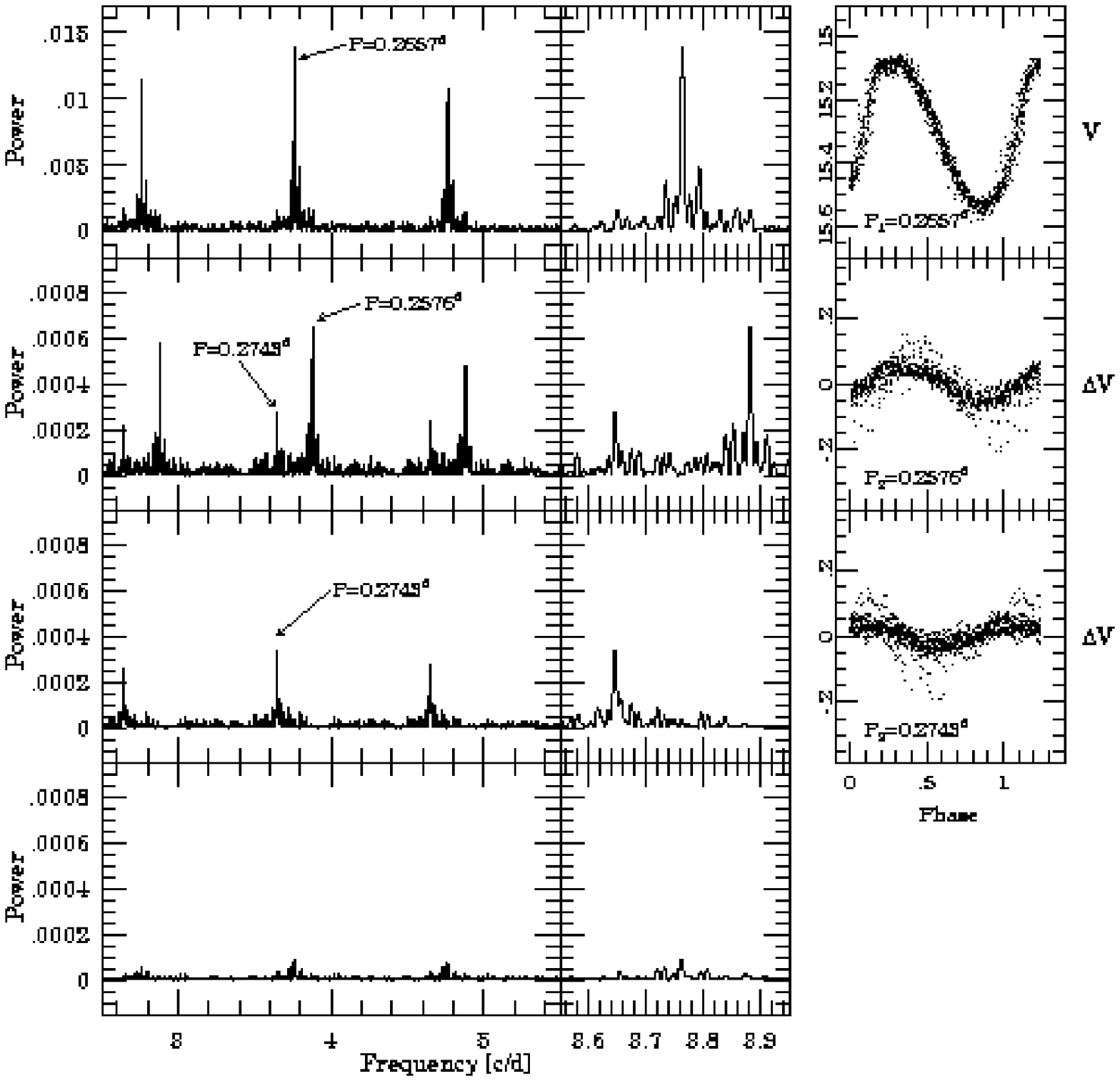}

\clearpage

\noindent {\bf Fig. 8} The power spectrum of RRab variable V18.
\includegraphics{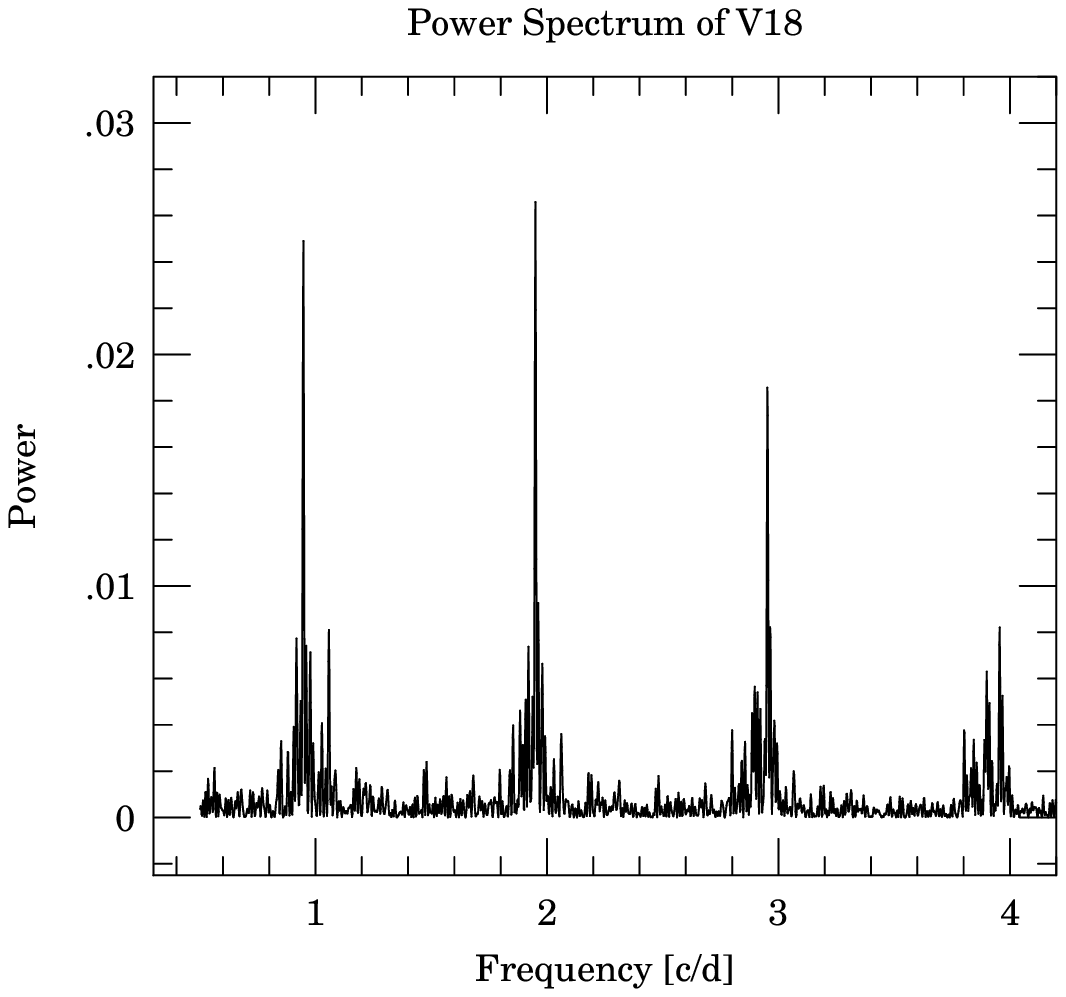}

\clearpage

\noindent {\bf Fig. 9} Histograms of the distribution in $V$ of the
HB stars in the color interval $0.0<B-V<0.85$ based on color magnitude
diagrams obtained by Alcaino and Liller (1986), Mazur et al. (1999),
Piotto et al. (1999) and in this work.
\includegraphics{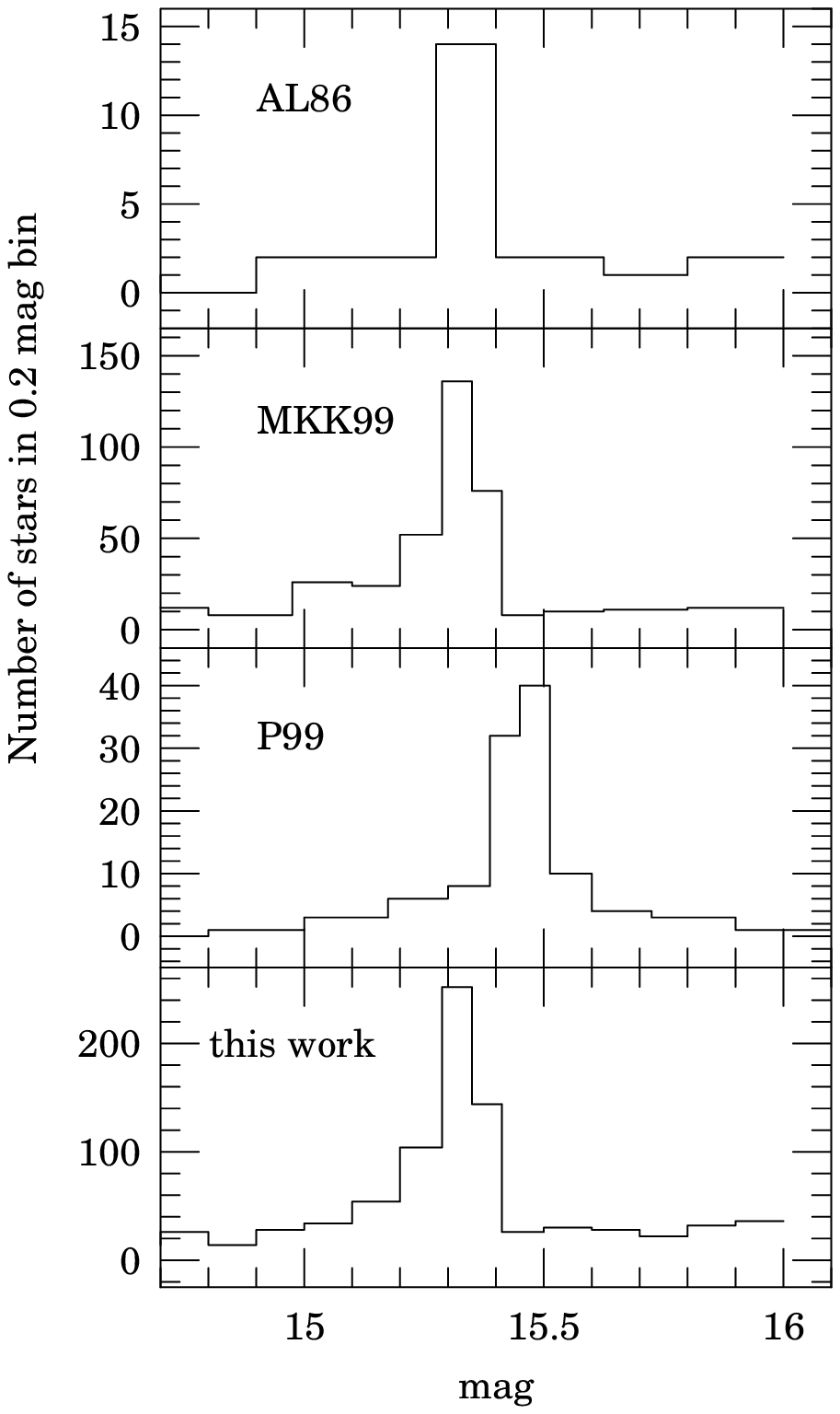}

\label{lastpage}

\end{document}